\documentclass[10pt,conference]{IEEEtran}
\usepackage{graphicx}
\usepackage{amssymb}
\usepackage{bm}

\begin{document}
\title{Layered Index-less Indexed Flash Codes for Improving Average Performance }

\author{\authorblockN{Riki Suzuki, \quad Tadashi Wadayama }
\authorblockA{$^\dag$ Department of Computer Science,  Nagoya Institute of Technology\\
Email: wadayama@nitech.ac.jp\\
}
} 

\maketitle
\begin{abstract}
In the present paper, a modification of the Index-less Indexed Flash Codes (ILIFC) for flash memory storage system is presented.
Although the ILIFC proposed by Mahdavifar et al. has 
excellent worst case performance, the ILIFC can be further improved in terms of the average case performance.
The proposed scheme, referred to as the {\em layered ILIFC},  is based on the ILIFC. 
However, the primary focus of the present study is the average case performance.
The main feature of the proposed scheme is the use of the layer-based index coding to represent indices of information bits. The layer index coding promotes the uniform use of cell levels, which leads to better average case performance.
Based on experiments, the proposed scheme achieves a larger average number of rewritings than the original ILIFC without loss of worst case performance. 
\end{abstract}

\section{Introduction}

%Flash memories used in solid state drives (SSD)
%are currently replacing hard disk drives (HDD) which have 
%been common storage devises  for digital equipments.

The emergence of studies by Jiang, Bohossian, and Bruck \cite{Jiang}\cite{Bruck} on flash codes 
gave rise to a new field of coding. They modeled a flash memory as a Write Asymmetric Memory (WAM)  
and proposed flash codes for improving the worst number of rewritings. 
These studies attracted the interest of a number of coding theorists 
and inspired subsequent research on flash codes.
In $2008$, Yaakobi, Vardy, Siegel, and Wolf presented flash codes based on the enhanced multidimensional construction \cite{EMC},
and they proposed a novel criterion, called {\em write deficiency},  for flash codes.

Most flash codes proposed thus far are designed to optimize the worst case performance, such as the write deficiency
\cite{Jiang}\cite{Bruck}\cite{EMC}\cite{ILIFC}.
On the other hand, recently, a flash code to improve 
the average number of rewritable bits was reported \cite{Average} \cite{Harvard} \cite{Kamabe}.
The {\em average number of rewritable bits} is the average number of allowable bit flips between consecutive erase operations.
The flash codes based on the Gray codes proposed by Finucane, Liu, and Mitzenmacher \cite{Average}
exhibit excellent average performance. They also presented a method for analyzing the average number of rewritable bits,
which is based on a Markov chain model constructed from the state diagram of the code and a probabilistic model for the rewriting process \cite{Average}.
In the lifetime of a flash memory, it is expected that poor average performance will result in early collapse of the cells. In this respect, flash codes should be designed to improve not only the worst case performance but also the average case performance.

The Index-less Indexed Flash Codes (ILIFC) proposed 
by Mahdavifar, Siegel, Vardy, Wolf, and Yaakobi \cite{ILIFC} achieve excellent worst case performance. The ILIFC has been proven to provide almost optimal write deficiency.
The prominent feature of the ILIFC is that a sub-block of cells represents both the value of an information bit and
the index of the bit. This feature leads to  simple encoding and decoding procedures.

In the present paper,  we present an improvement of the ILIFC
in terms of average case performance.
Since the ILIFC is designed based on the worst case performance, there is room for improving the average case performance without incorporating drastic changes in the original algorithm.

The main concept of the scheme proposed herein is the use of {\em layer-based index coding}. In the original ILIFC, if a sub-block represents an index, the index cannot be changed to another index until the next erase operation. If bit flips in the information vector occur according to a nonuniform distribution, the difference between cell levels tends to 
be large. It is not trivial to balance the cell levels if most of the sub-blocks have their own indices.
The layer-based index coding enables the encoder to change the index of a sub-block, and this feature provides flexibility for adjusting the differences between cell levels.

In the present paper, the {\em layered ILIFC}, which uses the layer-based index coding, will be proposed, and its average case performance will be evaluated through computer simulations and a Markov chain method designed for analyzing the average case performance of variants of the ILIFC.

\section{Preliminaries}

In this section, some basic assumptions on flash memories 
and a rewriting model are introduced according to \cite{Jiang}. 

\subsection{Flash memory}

A flash memory contains several cells that can store electrons.
The charge in a cell can be increased through the injection of electrons.
The process of increasing the cell level is referred to as {\em cell programming}.
A cell can represent $q$-ary values. For example, flash memories with $q = 2$ and $4$ cell levels have been realized in commercial products. The erase operation is the operation to remove the charge from cells in an {\em erase block}, which is a set of cells.
Note that the erase operation can only be applied to all of the cells in an erase block.
In other words, a cell cannot be erased individually.

There is a limitation in the number of rewritings for a cell. 
If the number of rewritings exceeds this limit, the cell tends to operate incorrectly.
Therefore, an appropriate coding for reducing the number of rewritings is required 
in order to lengthen the lifetime of a flash memory \cite{Jiang}\cite{ILIFC}.

\subsection{Rewriting model}

In this subsection, a simple rewriting model used throughout the present paper is introduced \cite{Jiang}\cite{ILIFC}. The {\em information vector} $(v_{1},v_{2},\cdots,v_{k})$ is a binary $k$-tuple, which is to be stored in an erase block. The initial state of the information vector is assumed to be $(v_{1},v_{2},\cdots,v_{k}) = (0,0,\cdots,0)$. The binary information in the information vector is written into cells through several rewriting processes.
It is assumed that a rewriting process occurs when any one bit in the information vector flips.

The {\em cell state vector} for an erase block is denoted by $(c_{1},c_{2},\cdots,c_{n})$, where 
$
c_{j} \in \{0,\cdots,q-1\} (j \in \{1,\ldots, n\}).
$
The symbol $c_j$ represents the state of the $j$th cell. 
The contents of the information vector are stored in the cell state vector.
The initial state of the cell state vector is also assumed to be  $(c_{1},c_{2},\cdots,c_{n}) = (0,0,\cdots,0)$.

Only two operations to change the state of a cell state vector are allowed, 
namely, cell programming and the erase operation.
Cell programming increases a cell level by one until the level reaches $q-1$. 
Suppose that there are two state vectors $\bm c = (c_{1},c_{2},\cdots,c_{n})$ and $\bm c' = (c'_{1},c'_{2},\cdots,c'_{n})$. If $c_{j}\geq  c'_{j}$ holds for any $j\in \{1,2,\cdots,n \}$, then $\bm c$ is said to be {\em higher} than $\bm c'$. These assumptions mean that the allowable state transitions of an erase block are a transition from a lower state to a higher state. An erase operation forces the cell state vector to be a zero vector.

A goal of the flash codes is to reduce the number of erase operations to be as small as possible,
under the assumption that the number of bit flips of information bits is fixed.

\section{Index-less indexed flash codes}

In this section, a brief introduction of the ILIFC as reported by \cite{ILIFC} is presented.
Although an encoding and decoding process of the ILIFC consists of multiple stages,
for simplicity, we herein focus only on the first stage.

Assume that an erase block contains $n$-cells and that these cells are divided into 
sub-blocks of the same size and any remaining cells. The size of a sub-block is assumed to be $k$ satisfying $n > k^2$, and $k \mbox{ mod } 2 = 0$.
The number of sub-blocks is given by $m = \lfloor n/k \rfloor$, and the number of remaining cells 
is $n - k m$. The length of the information vector is $k$. 
The key feature of the ILIFC is the representation of individual bits in the information vector.
Each sub-block in an erase block represents both an index and a value of an information bit.

Let us denote the state of an erase block as 
$
({\bm x_{1}}\mid{\bm x_{2}}\mid \cdots \mid{\bm x_{m}})
$
where $\bm x_{i} \in \{0,\ldots, q-1\}^k (i \in \{1,\ldots, m\})$ is the $i$th sub-block.
For a sub-block ${\bm x} = (c_{1},c_{2},\cdots,c_{k})$, 
the weight and the parity of ${\bm x}$ are defined as
$
{\rm wt}({\bm x}) = c_{1}+c_{2}+\cdots+c_{k},
$
and 
$
{\rm parity}({\bm x}) = {\rm wt}({\bm x})\ {\rm mod}\ 2,
$
respectively.

The following terminology is used throughout the present paper.
A state of a sub-block is said to be 
\begin{enumerate}
\item {\em full} if all cells in the sub-block have level $q-1$;
\item {\em empty} if all cells in the sub-block have level $0$;
\item {\em active } if the sub-block is neither full nor empty;
\item {\em live } if the sub-block is not live.
\end{enumerate}

\subsection{Index coding for the ILIFC}
A sub-block represents a single bit in the information vector.
The details are as follows. Let 
$
I: \{0,\ldots, q-1\}^k \rightarrow \{0,1,2, \ldots, k\}
$
be an {\em index map} that converts a state of a sub-block 
into an index of the information vector. In  \cite{ILIFC}, the index map is given by
\begin{equation}
I(\bm{x}) = 
\left\{
\begin{array}{l}
\arg \max_{i \in \{1,\ldots, k\}} [ c_i - c_{( i-2 \mbox{ \scriptsize mod } k)+1} ], \quad \bm x  \ne \bm 0 \\
0, \quad \mbox{otherwise},
\end{array}
\right.
\end{equation}
where ${\bm x} = (c_{1},c_{2},\cdots,c_{k})$. 
Note that $I(\bm{x}) = 0$ implies that the sub-block $\bm x$ has no index.
The {\em bit value map} for a sub-block: 
$
V: \{0,\ldots, q-1\}^k \rightarrow \{0, 1\}
$
is given simply as
$
V(\bm{x}) = \mbox{parity}(\bm x ).
$

Based on the definition presented above, a sub-block $\bm{x}_i$ can represent both 
an index $I(\bm{x}_i)$ and the bit value $V(\bm{x}_i)$.
At any moment during an encoding process (to be described later), the ILIFC encoder maintains the consistency as follows:
$
v_{I(\bm x_i)} = V(\bm x_i) \ \mbox{ if } I(\bm{x}_i)  \ne 0
$
for $i = 1,2,\ldots, k$, where $(v_1,\ldots, v_k)$ is the information vector.
A coding that achieves this consistency is referred to as an {\em index coding}.

\begin{figure}[tb]
\centering
\includegraphics[width=0.9 \linewidth]{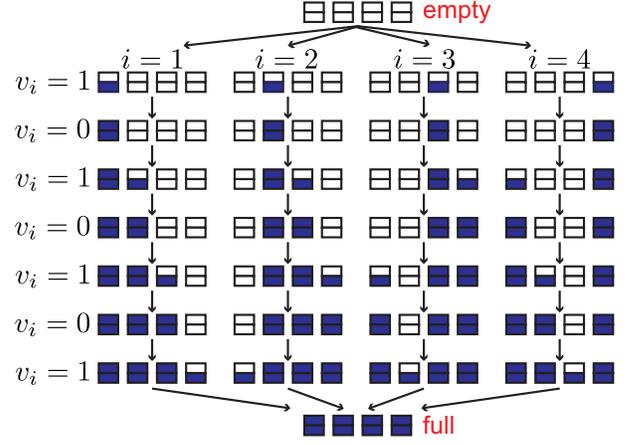}
\caption{Index coding for the ILIFC ($k = 4, q = 3$).
The variable $i$ expresses the value of the index. 
Whenever information bit $v_{i}$ turns, 
a state transition (an arrow) from a cell state vector to a higher cell state vector occurs.}

\label{fig:ILIFC_state}
\end{figure}

In the ILIFC, the index coding for a sub-block is carefully designed in order to avoid an early erase operation. Figure \ref{fig:ILIFC_state} illustrates the process for the case in which $k = 4$ and $q = 2$. In Fig. \ref{fig:ILIFC_state}, the variable $i$ corresponds to the index.
The position at which the level of the cells begins to rise corresponds to the index $i$ and is consistent with the definition of $I(\cdot)$.
Furthermore, a bit flip on $v_{i}$ induces an increment of a cell value 
that corresponds to the arrow shown in Fig. \ref{fig:ILIFC_state}.
Note also that the states of two active sub-blocks 
cannot have the same state if  the indices of two sub-blocks are different.

\subsection{Pseudo-code for the ILIFC encoder and decoder}

The decoding and encoding algorithms for the ILIFC are shown in the form of a pseudo-code according to \cite{ILIFC}.

%\begin{screen}
\noindent
{\bf Decoding\ map\ } \\
$(v_{1},v_{2},\cdots,v_{k}) = (0,0,\cdots,0)$;\\
{\bf for\ ($j = 1$;$j \leq  m$;$j = j+1$)}\\
{\bf if\ (active($\bm{x}_{j}$))}\\
{\bf \{ $i = $ read{\_}index($\bm{x}_{j}$);$v_{i} = $ parity(${\bm{x}}_{j}$);\} }
%\end{screen}
\vspace{0.5cm}

%\begin{screen}
\noindent{\bf Encoding\ map\ } \\
${\bm y}=({\bm{y}}_{1}\mid {\bm{y}}_{2}\mid \cdots \mid {\bm{y}}_{m})=({\bm{x}}_{1}\mid {\bm{x}}_{2}\mid \cdots \mid {\bm{x}}_{m})$;\\
{\bf for\ ($j = 1$;$j \leq m$;$j = j+1$) \ \{ } \\
\ \ {\bf if\ (active(${\bm{x}}_{j}$) $\wedge$ (read{\_}index(${\bm{x}}_{j}$) $== i$))}\\
\ \ {\bf \{write(${\bm{y}}_{j}$);return(${\bm y}$);\} }\\
{\bf \} } \\ \\
{\bf for\ ($j = 1$;$j \leq  m$;$j = j+1$)}\\
\ \ {\bf if\ (empty(${\bm{x}}_{j}$))\\
\ \{write{\_}new($i, {\bm {y}}_{j}$);return(${\bm y}$);\} } \\
{\bf return\ E};
%\end{screen}

The symbol {\bf E} represents the erase operation.
The details of the functions, such as {\bf active}$(\cdot)$, can be found in \cite{ILIFC}.

\section{Layered ILIFC for improving average case performance}

In this section, we propose a modified ILIFC to improve the average case performance.
A significant difference between the original ILIFC and the modified ILIFC is the index coding 
for a sub-block. The modified ILIFC introduces the concept of the {\em layer} in its
index coding. The layer in a sub-block enables us to reset a sub-block, which means
that an index embedded in a sub-block can be changed to another index. This flexibility 
promotes the uniform use of cell levels in an erase block and leads to an improvement in 
the average case performance.

\subsection{Outline of Layered ILIFC}

In the following, for simplicity, we assume that an erase block contains exactly $n = k^2$ cells.

In the encoding process of the original ILIFC,
the index of a sub-block is fixed until the next erase operation occurs.
Thus, if several bits in the information vector are frequently rewritten,  
the sub-blocks corresponding to these bits tend to become full in early phase.
In other words, local rewritings in the information vector induce an imbalance of cell levels between the sub-blocks. This imbalance tends to cause an early erase operation.
In order to overcome this problem, maintaining the balance of cell levels for any rewriting sequence is crucially important.

In the modified ILIFC, the cells in a sub-block are programmed from the bottom layer to the top layer. If a layer is filled (or programmed),  we may alter the index of the sub-block.
In the following, we refer to the proposed modified scheme as the {\em layered-ILIFC} (L-ILIFC).

\subsection{Index coding for Layered ILIFC}

In the following, we assume that $k$ is an even number.
Suppose that a cell $c_i$ has the value $l \in \{0,\ldots, q-1\}$. 
In such a case, the cell $c_i$ is said to be in layer $l$. In other words,
cell levels are divided into $q$-layers.

Next, we introduce some additional considerations regarding the state of a sub-block.
If all of the cells in a sub-block belong to the same layer $l \in \{0,1,\ldots,q-2\}$, then the sub-block is said to be {\em clear}. 

\begin{figure}[tb]
\centering
\includegraphics[width=0.9 \linewidth]{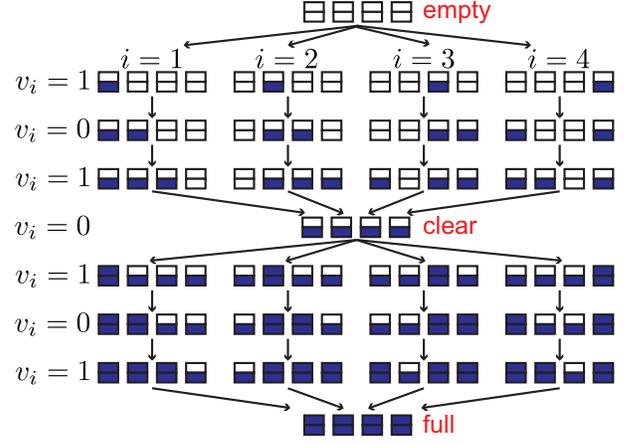}
\caption{Index coding for the layered ILIFC ($k = 4, q = 3$);  
The variable $i$ expresses the value of the index. 
Whenever information bit $v_{i}$ turns, 
a state transition (an arrow) from a cell state vector to a higher cell state vector occurs.}
\label{fig:L-ILIFC_state}
\end{figure}

We first explain the concept of index coding using the following example.
The details of the index coding will be presented in Subsection \ref{sec:dofL}.
Figure \ref{fig:L-ILIFC_state} presents the details of the index coding for $k = 4$ and $q = 2$.
Initially, the cell is empty (empty boxes at the top in Fig. \ref{fig:L-ILIFC_state}). A single bit change in the information vector corresponds to the arrows emerging from the empty state. An index $i$ (from 1 to 4) is written in a sub-block by incrementing the value of the $i$th cell, and
the parity of the sub-block becomes one.

It is readily observed that two more changes in the information vector transforms the state of 
the sub-block into the clear state. Namely, all of the cells belong to the same layer.
A clear sub-block has no index. Note that we can write any index into a clear sub-block, as shown in Fig. \ref{fig:L-ILIFC_state}.

The index map for this index coding is, thus, given by
\begin{equation}
I'(\bm{x}) = 
\left\{
\begin{array}{l}
0, \forall i,j \in \{1,\ldots, k\}, c_i = c_k, \mbox{(i.e., $\bm x$ is clear)} \\
\arg \max_{i \in \{1,\ldots, k\}} [ c_i - c_{( i-2 \mbox{ \scriptsize mod } k)+1} ], \mbox{otherwise}.
\end{array}
\right.
\end{equation}

Due to the assumption that $k$ is even,  any clear sub-block has a value of zero.
This is because the sum $c_1+\ldots + c_k$ becomes an integer multiple of $k$.
Therefore, there is no problemfor a sub-block to forget both its index and the value when it becomes a clear sub-block.

\subsection{Details of the layered ILIFC}
\label{sec:dofL}
The pseudo-code of the encoding map of the layered ILIFC is shown below.

%\begin{screen}
\noindent
{\bf Encoding\ map\ } \\
${\bm y}=({\bm{y}}_{1}\mid {\bm{y}}_{2}\mid \cdots \mid {\bm{y}}_{m})=({\bm{x}}_{1}\mid {\bm{x}}_{2}\mid \cdots \mid {\bm{x}}_{m})$;\\
{\bf for\ ($j = 1$; $j \leq m$; $j = j+1$) \{ } \\
\ \ {\bf if\ (($\lnot$ clear(${\bm{x}}_{j}$)) $\wedge$ (read{\_}index2(${\bm{x}}_{j}$) $== i$))} \\
\ \ {\bf \{ write2(${\bm{y}}_{i}$); return(${\bm y}$);\} } \\
{\bf \} }\\ \\ 
{\bf for\ ($l=0$; $l \leq q-1$; $l=l+1$) \{ } \\
\ \ {\bf for\ ($j=1$; $j \leq m$; $j = j+1$) \{ } \\
\ \ \ \ {\bf if\ (clear(${\bm {x}}_{j}$) $\wedge$ (read{\_}layer(${\bm x}_{j}$)$== l$))} \\
\ \ \ {\bf \{write{\_}new2($i,{\bm y}_{j}$); return(${\bm y}$);\} } \\
{\bf \ \ \} } \\
{\bf \} } \\ 
{\bf return\ E};
%\end{screen}

The clear sub-blocks in an erase block  can be considered to be a resource for achieving flexibility.
Assume that the encoder must write the index value $i$ in a sub-block.
If there is no active sub-block with the index value $i$, then the encoder searches for a clear sub-block in the lowest layer. Note that the encoder of the original ILIFC can choose a sub-block to write the index value $i$ only from among the empty sub-blocks. In other words, the existence of clear sub-blocks and this search method reduces the nonuniform use of the cell levels.

The details of the functions used in the pseudo-code are given as follows.
The function {\bf clear}(${\bm{x}}$) returns true if ${\bm{x}}$ is clear; otherwise the function returns false.
The function {\bf read\_layer}(${\bm {x}}$) returns the layer index 
of ${\bm {x}}$ if ${\bm {x}}$ is clear; otherwise the function returns $-1$.

The function {\bf read\_index2}(${\bm {x}}$) is defined as
\[
\mbox{{\bf read\_index2}(${\bm {x}}$)} = I'(\bm x).
\]

The function {\bf write\_new2}($i,{\bm x}$) is a function to write the index value $i$ into the clear sub-block ${\bm x}$. This function simply increments the value of $c_i $ by one.
The function {\bf write2}(${\bm {x}}$) changes the value of the active sub-block ${\bm {x}}$ by
incrementing 
$
c_{(i+{\rm wt}({\bm {x}}) ) \bmod k}
$ where ${\bm x}=(c_1,\ldots, c_k)$.
This rule of the change in the value of a sub-block corresponds to the arrows in Fig. \ref{fig:L-ILIFC_state}.

\subsection{Worst case performance}

The worst number of possible rewritings between a consecutive erase operation for 
the proposed scheme is same as that for the original ILIFC.
This is because both schemes share the same worst case events.
A worst case event occurs when 
the cell vector contains one full sub-block and $k-1$ active sub-blocks of cell level 1.

\section{Markov chain method }
\label{sec:MCM}
In this section, we briefly review the Markov chain method for average case analysis \cite{Average}
and then apply the Markov chain method to the analysis on the ILIFC and the Layered ILIFC.

The concept of the Markov chain method as follows. 
We first construct a state transition diagram for a flash code to be analyzed.
A state of the diagram is an allowable state of the cell state vector. Each edge (i.e., state transition)
has its own probability, which is determined based on the probabilistic model of 
rewriting sequences for the information vector. 
The state transition diagram naturally defines a Markov chain corresponding the pair of
the flash code and the probabilistic model.

The steady state probability of the Markov chain can be obtained by solving a simultaneous  linear equation system 
defined by the transition probability matrix.  
The average number of possible rewritings is directly derived from the steady state probability.

\begin{figure}[tb]
\centering
\includegraphics[width=0.8 \linewidth]{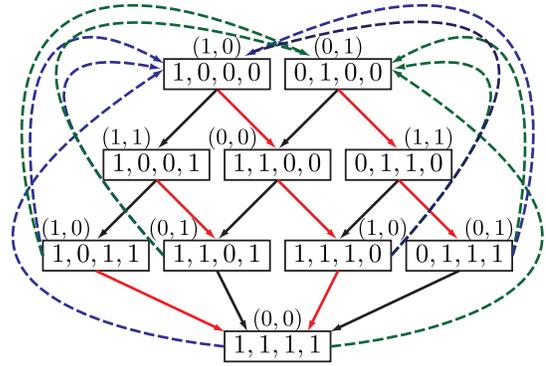}
\caption{An example of the Markov chain for ILIFC ($n = 4, k = 2, m = 2, q = 2$)}
\label{fig:Markov_chain}
\end{figure}

Figure \ref{fig:Markov_chain} illustrates the state transition diagram of
the ILIFC with the parameters $n = 4, k = 2$, and $q = 2$. The binary 4-tuples in the boxes are the states of the cell state vector ($n = 4$), and the binary 2-tuples in parentheses represent the states of the information vector.
The arcs connecting the boxes denote possible state transitions.
The dotted arcs correspond to the state transitions that induce an erase operation.

The average probability of the occurrence of an erase operation is given by the
sum of the probabilities of the state transitions corresponding to the dotted arcs.
This probability can be calculated from the steady state probability.
The average number of possible rewritings is given as the inverse of this probability \cite{Average}.

If the state space of the ILIFC or the Layered ILIFC is not so large, then this Markov chain method is useful to obtain 
an accurate estimate. Table \ref{tb:bp2} compares the average probability of an occurrence of an erase operation
obtained through computer simulation and the Markov chain method.  We have confirmed that 
the values obtained by the Markov method and the computer simulations are in reasonable agreement.
\begin{table}[tb]
\caption{Average probability of occurrence of erase operations of the layered ILIFC ($n = 4, k = 2, q = 4$).}
\label{tb:bp2}
\begin{center}
\begin{tabular}{lll}
\hline
 & $P_{1}=1/2$ & $P_{1}=1/5$ \\ \hline
Computer simulation  & $0.091006$ & $0.095430$  \\ \hline
Markov chain  & $0.091006$ & $0.095431$  \\
\hline
\end{tabular} \\
\end{center}
{\scriptsize The probability $P_1$ is the probability such that the first bit of the information bit is flipped.}
\end{table}

\section{Results of computer experiments}

We performed computer experiments to compare the Layered  ILIFC and the original ILIFC
in terms of the number of possible rewritings between two consecutive erase operations.

The assumptions made in the simulations are as follows.
In unit time, only one bit in the information vector flips, which induces a state change in the cell state vector.
The flipped bit is chosen according to the uniform distribution from 1 to $k$.
The same pseudo-random sequences are used for both schemes in order to realize a fair comparison. In the simulation, the information vector and the cell state vector are initialized after a block erase operation.

%\subsection{Comparison of the number of rewritings between consecutive erase operations}

Figure \ref{fig:ILIFCvsLILIFC,k=4,q=8} presents histograms for the number of 
rewritings between consecutive erase operations.
The horizontal axis represents the number of rewritings, and the vertical axis denotes
the frequency of the number of rewritings observed in the experiment.
In an experiment, rewriting operations are iterated until the number of erase operations becomes $10^4$-times. The parameter settings are as follows. The number of cells in an erase block is $n = 16$, and each cell can take a value from $0$ to $7$ (i.e., $q = 8$). The length of the information vector is $k = 4$, and thus $m = 4$.

It is readily observed that the histogram curve of the layered ILIFC indicates a desirable behavior of the layered ILIFC.
If the number of rewritings is not so large (i.e., smaller than 94), then the layered ILIFC provides a much smaller frequency than the ILIFC.
The figure also includes the average number of rewritings. The layered ILIFC yields 100.89 rewritings, and the original ILIFC yields 93.65 rewritings. 
On average, approximately seven more rewritings are possible with the layered ILIFC.
\begin{figure}[tb]
\centering
\includegraphics[width=1 \linewidth]{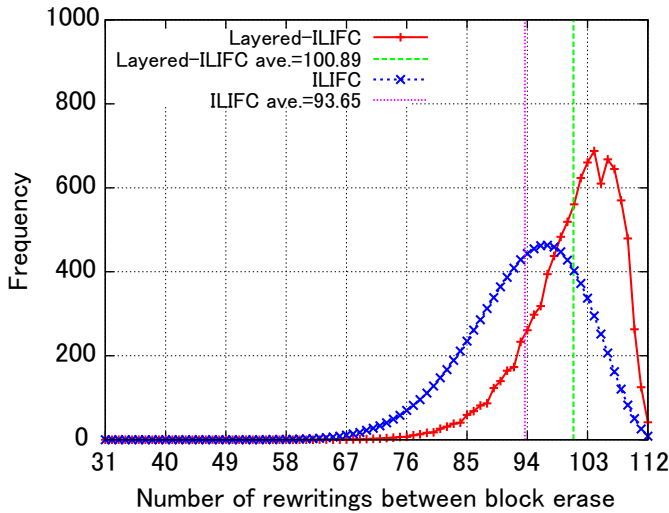}
\caption{Frequency of the number of rewritings between the block elimination of ILIFC and layered ILIFC ($n = 16, k = 4, m = 4, q = 8$).}
\label{fig:ILIFCvsLILIFC,k=4,q=8}
\end{figure}

%\subsection{Analysis of the average number of rewritings based on the Markov chain method}

The Markov chain method presented in Section \ref{sec:MCM} provides accurate values of 
the expected number of rewritings for the layered ILFC and the ILIFC.

\begin{figure}[tb]
\begin{center}
\includegraphics[width=1 \linewidth]{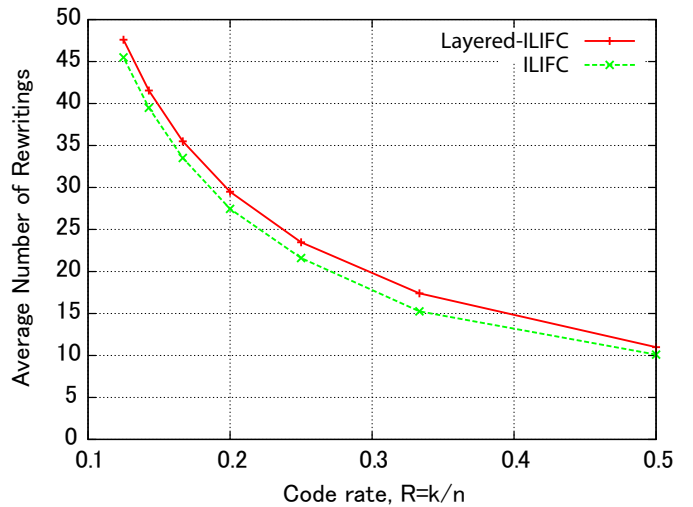}
\end{center}
\caption{Average number of rewritings between consecutive erase operations of ILIFC and layered ILIFC ($k = 2, q = 4$).}
\label{fig:IvsLNAvek=2q=4}
\end{figure}

Figure \ref{fig:IvsLNAvek=2q=4} shows the trade-off curves between the code rate versus the average number of rewritings.
The code rate is defined as $R = k/n$. The parameters $k = 2$ and $n = 4$ are assumed.
The layered ILIFC provides a better trade-off  compared with the original ILIFC.
This indicates that the proposed scheme has better average case performance than that of the original scheme.

\section{Conclusive summary}

The original ILIFC is an excellent flash code having near optimal write deficiency.
In the present paper, we revealed that a simple modification of the index coding promotes the uniform use of the cell level. In addition, we demonstrated that the average performance of the ILIFC can be further improved without loss of worst case performance.
The Markov chain method for the ILIFC may be a useful tool for optimizing the detail of the algorithm in terms of average performance.

\end{document}